\documentclass[12pt]{iopart}

\usepackage{graphicx} 
\begin{document}

\title{Elliptic flow from event-by-event hydrodynamics}

\author{\underline{H.~Holopainen}$^{1,2}$, H.~Niemi$^3$ and K.~J.~Eskola$^{1,2}$}

\address{$^1$ Department of Physics, P.O.Box 35, FI-40014 University of Jyv\"askyl\"a, Finland}
\address{$^2$ Helsinki Institute of Physics, P.O.Box 64, FI-00014 University of Helsinki, Finland}
\address{$^3$ Frankfurt Institute for Advanced Studies, Ruth-Moufang-Str. 1, D-60438 Frankfurt am Main, Germany}
\eads{\mailto{hannu.l.holopainen@jyu.fi}} 
\begin{abstract}
We present an event-by-event hydrodynamical framework which takes into account the initial
density fluctuations arising from a Monte Carlo Glauber model. The elliptic flow is
calculated with the event plane method and a one-to-one comparison with the measured
event plane $v_2$ is made. Both the centrality- and $p_T$-dependence of the $v_2$ are
remarkably well reproduced. We also find that  the participant plane is a quite good
approximation for the event plane.
\end{abstract}


\section{Introduction}
\label{sec: intro}
Hydrodynamical models using smooth initial conditions cannot reproduce
the centrality dependence of the elliptic flow coefficients $v_2$,
measured using the event plane method (see e.g. Fig.~7.5 in \cite{Niemi:2008zz}).
To study this more carefully we present here an event-by-event hydrodynamical 
framework introduced in Ref.~\cite{Holopainen:2010gz} and make a one-to-one 
comparison with the experimental event plane results for the elliptic flow. 
Also the $v_2$ with respect to the participant plane is considered.

\section{Event-by-event hydrodynamics framework}
\label{sec: ebye hydro}

The initial state is obtained from a Monte Carlo Glauber model. First, nucleons
are randomly distributed into the nuclei using a standard Woods-Saxon potential. Then
the impact parameter $b$ is sampled from a distribution $dN/db \sim b$. A nucleon $i$
and a nucleon $j$ from different nuclei collide if their transverse locations are
close enough,
\begin{equation}
  (x_i - x_j)^2 + (y_i - y_j)^2 \le \frac{\sigma_{NN}}{\pi},
\end{equation}
where $\sigma_{NN}$ is the inelastic nucleon-nucleon cross section. For
$\sqrt{s_{NN}} = 200$~GeV we use $\sigma_{NN} = 42$~mb. The impact parameter defines
the reaction plane (RP) and the participant plane (PP) is obtained from the participant
configuration by maximizing the eccentricity. Centrality classes are defined using the
number of participants $N_{\rm{part}}$ and slicing their distribution into $N_{\rm{part}}$
intervals as shown in Ref.~\cite{Holopainen:2010gz}.

\begin{figure}[t]
  \centering
  \includegraphics[height=7.0cm]{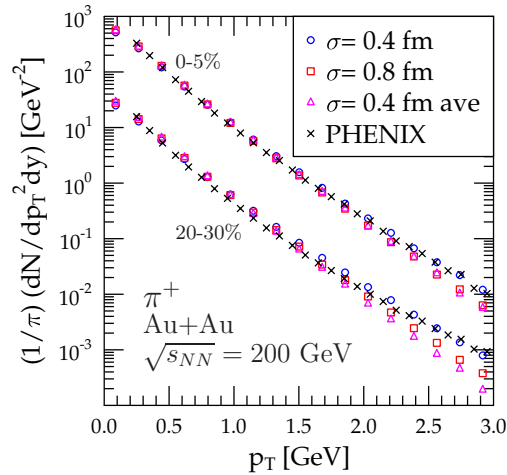}
  \caption{The transverse momentum spectra of positively charged pions in Au+Au collisions at
    $\sqrt{s_{NN}} = 200$~GeV. The event-by-event hydrodynamical calculations with two
    different fluctuation size parameters and the smooth initial state calculations
    are shown at two centralities. The data are from the PHENIX Collaboration
    \cite{Adler:2003cb}. From \cite{Holopainen:2010gz}.}
  \label{fig: pion spectra}
\end{figure}

The initial energy density $\epsilon$ in the transverse plane is obtained by distributing
energy around the participants using a 2-dimensional Gaussian smearing,
\begin{equation}
  \epsilon (x,y) = \frac{K}{2\pi \sigma^2} \sum_{i=1}^{N_{\rm{part}}} \exp\Big( -\frac{(x-x_i)^2+(y-y_i)^2}{2\sigma^2} \Big),
  \label{eq:eps}
\end{equation}
where $K$ is a overall normalization factor and $\sigma$ is a free parameter, which
controls the width of the Gaussian. The overall normalization as well as the initial
time $\tau_0=0.17$~fm are fixed from the EKRT model \cite{Eskola:1999fc}. 

For each event, we solve the ideal hydrodynamical equations $\partial_\mu T^{\mu\nu} = 0$ assuming
longitudinal boost-invariance and zero net-baryon density.
We further need to specify an equation of state (EoS) $P=P(\epsilon)$ to close the set of equations. 
Our choice is the EoS from Ref.~\cite{Laine:2006cp}. The freeze-out is assumed to happen at a constant
temperature $T_F = 160$~MeV, and the thermal spectra are calculated using the Cooper-Frye
method. Hadrons are sampled from the calculated spectra and they are given to
PYTHIA 6.4 \cite{Sjostrand:2006za}, which does all the strong and electromagnetic
resonance decays.
 
In order to make a one-to-one comparison with the experiments, the elliptic flow is
calculated with the event plane (EP) method. The event flow vector for the second
harmonic is
\begin{equation}
  Q_2 = \sum_i ( p_{Ti} \cos(2\phi_i),  p_{Ti} \sin(2\phi_i) ),
\end{equation}
where we sum over all particles, and the event plane angle $\psi_2$ is obtained from this as
\begin{equation}
  \psi_2 = \frac{ {\rm arctan} ( Q_{2,y} / Q_{2,x} ) }{2}.
\end{equation}
Since we have only a finite number number of particles in each event, the event plane
fluctuates around the true event plane, which would be the one determined from
infinitely many particles. The correction $\mathcal{R}$ from these fluctuations is
calculated with the 2-subevent method \cite{Poskanzer:1998yz}. The final event plane
elliptic flow is thus
\begin{equation}
  v_2\{ {\rm EP} \} = \langle \cos (2(\phi_i - \psi_2)) \rangle /\mathcal{R}.
\end{equation}

\begin{figure}[th]
  \begin{center}
  \includegraphics[height=10.5cm]{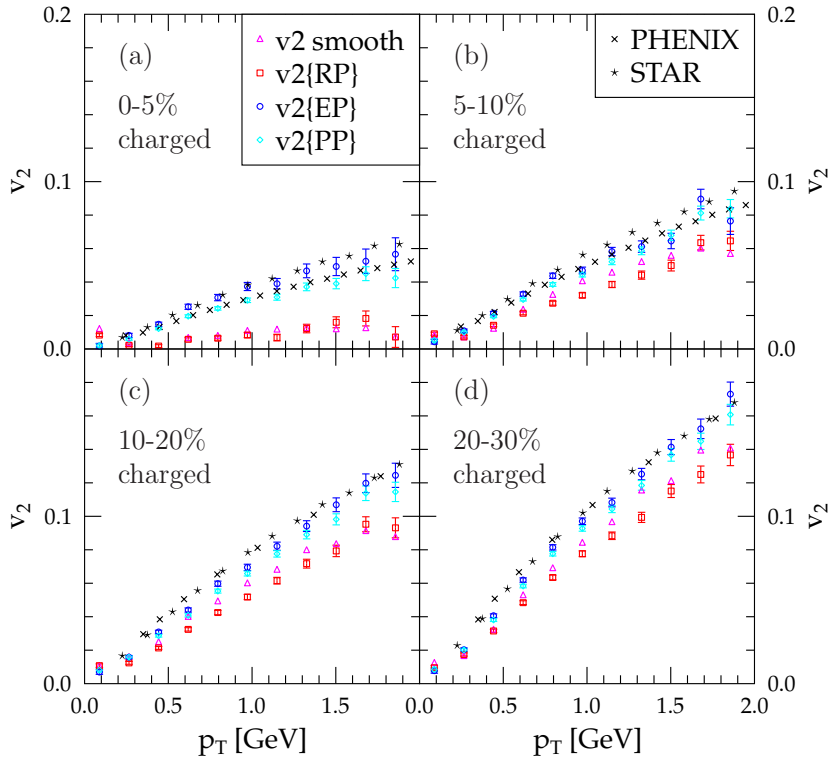}
  \end{center}
  \caption{The elliptic flow of charged hadrons in Au+Au collisions at
    $\sqrt{s_{NN}} = 200$~GeV. The event-by-event calculations with respect
    to different reference planes and the calculations with smooth initial states are
    from \cite{Holopainen:2010gz}. The data are from the PHENIX and STAR Collaborations
    \cite{Afanasiev:2009wq,Adare:2010ux,Adams:2004bi}.  }
  \label{fig: v2 charged}
\end{figure}

\section{Results}
\label{sec: results}

For each centrality class we have made 500 hydro runs and to increase the event statistics
we have sampled each thermal hadron spectra 20 times. The smooth initial states are obtained
by taking an average over 20 000 initial energy density profiles. The obtained transverse
momentum spectra of pions are plotted in Fig.~\ref{fig: pion spectra}. The density
fluctuations increase the number of high-$p_T$ particles since the initial pressure
gradients are larger. However, if the fluctuation size parameter $\sigma$ is sufficiently 
large, the difference to the smooth result is small.

In Fig.~\ref{fig: v2 charged} we have plotted the elliptic flow with respect to the
different reference planes. The EP results agree very well with the measured data
at all centralities shown in the figure. If we use PP instead of EP, the elliptic
flow is a little bit smaller, but PP seems to be a quite good approximation for the
EP. The $v_2\{\rm RP\}$ is generally smaller than $v_2\{\rm EP\}$, but it is very
close to the smooth result. Thus the fluctuations alone do not generate more elliptic
flow, but the reference plane definition is very important.

\begin{figure}[t]
  \centering
  \includegraphics[height=6.5cm]{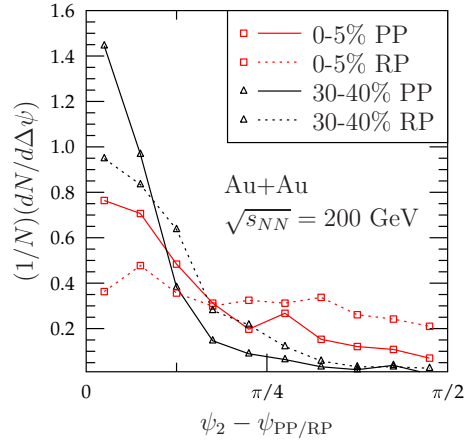}
  \caption{The correlation of the event plane with the participant plane and the reaction
    plane in Au+Au collisions at $\sqrt{s_{NN}} = 200$~GeV. From \cite{Holopainen:2010gz}.}
  \label{fig: ep vs rp}
\end{figure}

In Fig.~\ref{fig: ep vs rp} is shown the correlation of the EP with the PP, and with
the RP. Note that the trivial fluctuations of $\psi_2$ around the true EP are included
in the figure. The correlation between the EP and the PP is, as expected, stronger than
that between the EP and the RP.

\ack

This work was supported by the Academy of Finland (project 133005), the Finnish Graduate
School of Particle and Nuclear Physics, the
Magnus Ehrnrooth Foundation and the Extreme Matter Institute (EMMI). We acknowledge
CSC -- IT Center for Science in Espoo, Finland, for the allocation of computational
resources.

\end{document}